\documentclass{spie}

\usepackage{amsmath,amsfonts,amssymb}
\usepackage{graphicx}
\usepackage[colorlinks=true, allcolors=blue]{hyperref}
\usepackage{sidecap}
\usepackage[dvipsnames]{xcolor}
\usepackage{multirow}

\def\deg{\hbox{$^\circ$}}

\def\lae{\mathrel{\raise .4ex\hbox{\rlap{$<$}\lower 1.2ex\hbox{$\sim$}}}}
\def\gae{\mathrel{\raise .4ex\hbox{\rlap{$>$}\lower 1.2ex\hbox{$\sim$}}}}

\title{Characterization of X-ray Detectors in the MIT X-ray Polarimetry Beamline}

\author[a]{Sarah N.\ T.\ Heine}
\author[a]{Herman L.\ Marshall}
\author[a]{Benjamin Schneider}
\author[a]{Beverly LaMarr}
\author[a]{Nithya Kothnur}
\author[a]{Alan Garner}

\affil[a]{MIT Kavli Institute, Cambridge, MA, USA}

\authorinfo{Further author information: (Send correspondence to S.N.T.H.)\\S.N.T.H..: saraht@mit.edu,
Telephone: 1 617 258 8819} 

\begin{document}

\maketitle

\begin{abstract}

The MIT X-ray Polarimetry Beamline is a facility that we developed for testing components for possible use in X-ray polarimetry.  Over the past few years, we have demonstrated that the X-ray source can generate nearly 100\% polarized X-rays at various energies from 183 eV (Boron K$\alpha$) to 705 eV (Fe L$\alpha$) using a laterally graded multilayer coated mirror (LGML) oriented at 45\deg\ to the source.  The position angle of the polarization can be rotated through a range of roughly 150\deg.  In a downstream chamber, we can orient a Princeton Instruments MTE1300B CCD camera to observe the polarized light either directly or after reflection at 45\deg\ by a second LGML.  In support of the REDSoX Polarimeter project, we have tested four other detectors by directly comparing them to the PI camera.  Two were CCD cameras: a Raptor Eagle XV and a CCID94 produced by MIT Lincoln Laboratories, and two had sCMOS sensors: the Sydor Wraith with a GSENSE 400BSI sensor and a custom Sony IMX290 sensor.  We will show results comparing quantum efficiencies and event images in the soft X-ray band.

\end{abstract}


\section{REDSoX Detector Requirements}
\begin{figure}
   \centering
    \includegraphics[width=17cm]{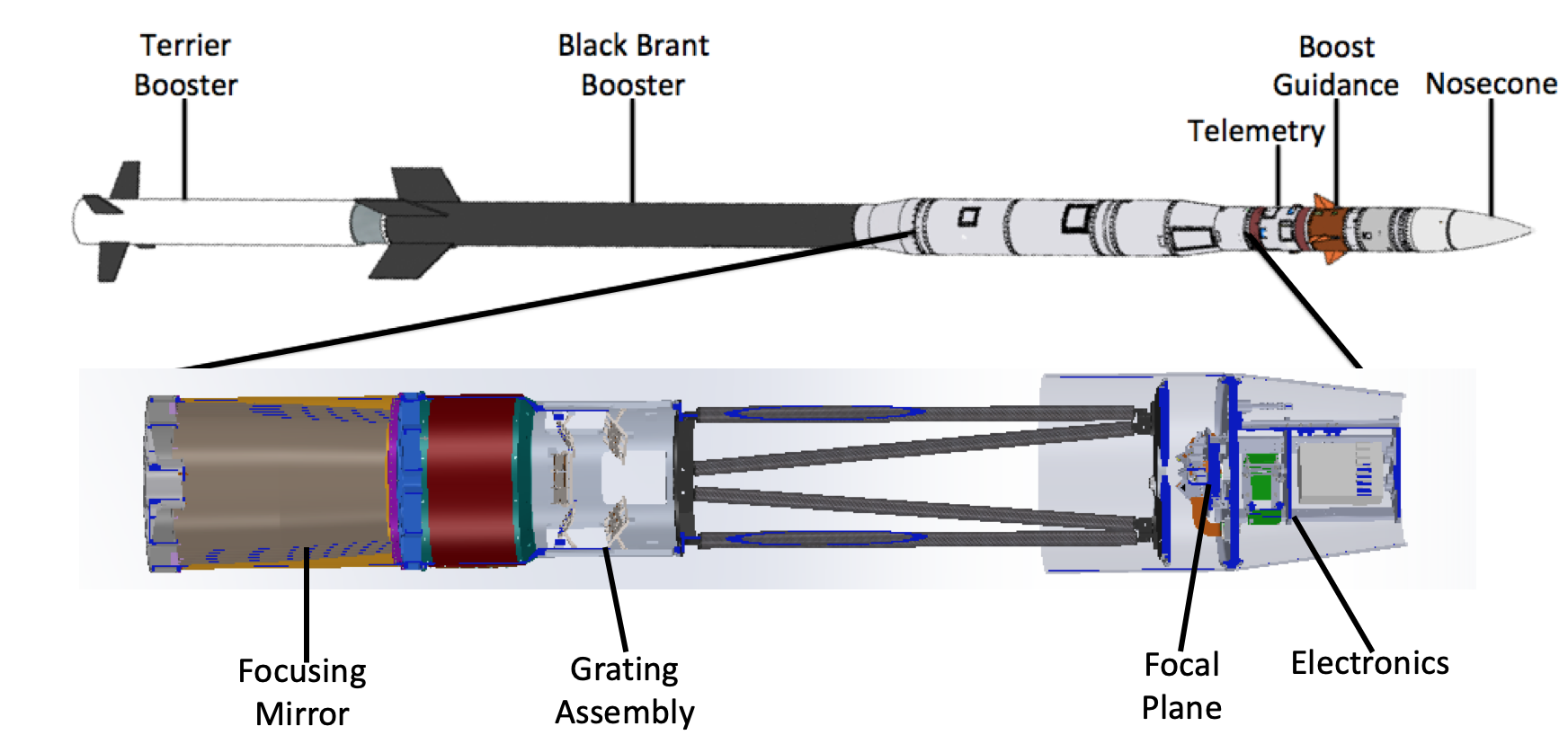}
 \caption{CAD representations of the full rocket (top) and the REDSoX payload (bottom).  Major systems and components are labelled.  The design is expected to be finalized by the time of the critical design review in 2025.}
\label{fig:REDSoXOverview}
\end{figure}

\begin{figure}
   \centering
    \includegraphics[width=9cm]{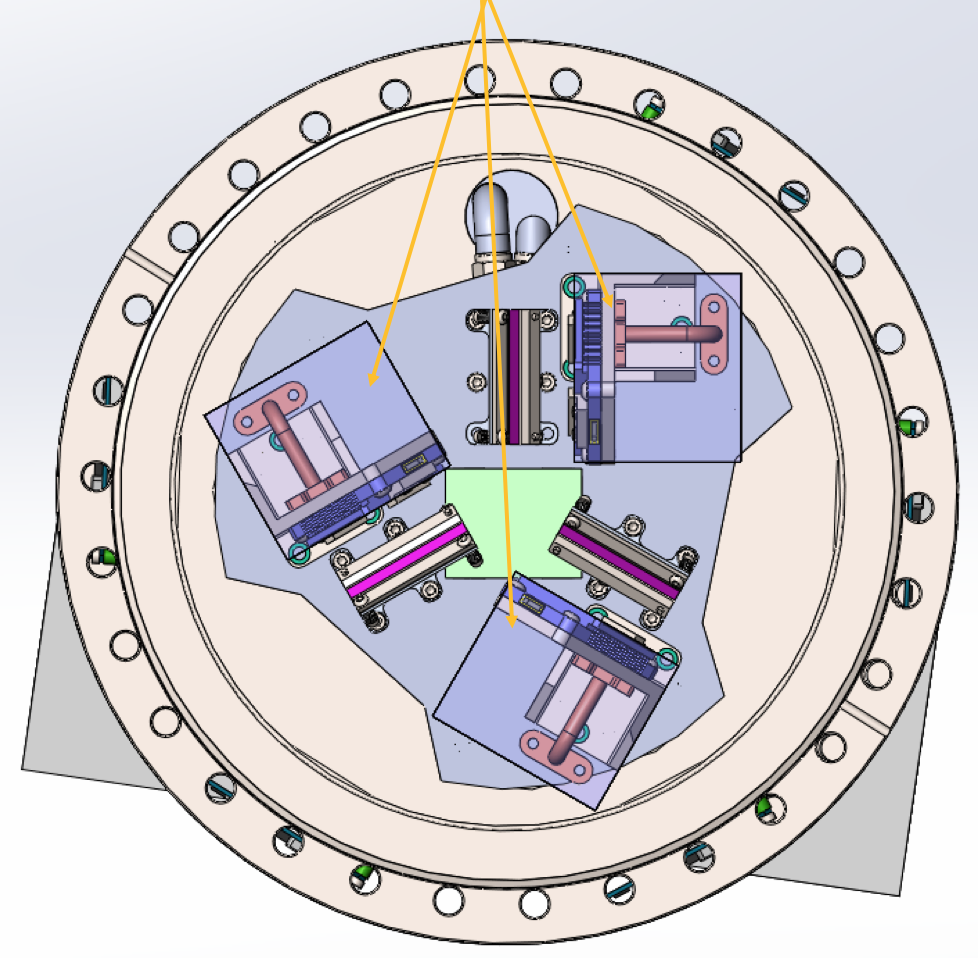}
 \caption{A CAD rendering of part of the REDSoX focal plane with boxes highlighting the available physical envelopes for the three polarimetry cameras  These boxes are highlighted with yellow arrows to guide the eye.  The imaging camera is not shown, for clarity.}
\label{fig:CameraEnvelopes}
\end{figure}

The Rocket Experiment Demonstration of a Soft X-ray (REDSoX) polarimeter is a NASA funded sounding rocket mission planned for launch in the summer of 2027 \cite{redsoxjatis,redsox24}.  The mission team completed a preliminary design review in March 2024 and we are currently working to finalize designs before the critical design review, currently planned for sometime in the first quarter of 2025.  REDSoX is a soft X-ray polarimeter operating in the 200 to 400 eV band.  CAD drawings of the rocket and the REDSoX payload are shown in Figure \ref{fig:REDSoXOverview}.  A mirror assembly from Marshall Space Flight Center will focus incoming X-ray light with a focal length of 2.5 m.  Critical Angle Transmission (CAT) gratings \cite{2022HeilmannCAT} disperse X-rays onto three Laterally Graded Multilayer Mirrors (LGMLs) oriented at 120\deg\ with respect to each other.  Each LGML reflects X-rays polarized parallel to its surface onto a detector.  The layout of the focal plane is shown in Figure \ref{fig:CameraEnvelopes}.

The requirements for these detectors include mild timing (1-10 second frame time) with a 1-3 second or shorter readout, and spectral resolution (full width at half maximum of less than 200 eV -- enough to reduce background).  The chip needs to be able to capture the spatial extent of the reflected spectrum, which is about 37 mm.  This can be along the diagonal of a rectangular chip.  The readout noise must be below 5 electrons to allow clear identification of 200 eV X-rays above the noise floor.  We also require a quantum efficiency of above 40\% across the bandpass, with a goal of 70\%.  We also utilize an imaging detector in the center of the focal plane for source acquisition and fine pointing.  This detector requires a faster readout but has more modest noise requirements.  Ideally this detector would be identical in design to the polarimetry detectors to allow for simplicity in readout design and interchangeability of cameras and flight spares.

One of the more difficult requirements to meet is that detectors and their packaging also need to fit within our focal plane, which is crowded with optical elements.  The envelope available for the detectors is shown in Figure \ref{fig:CameraEnvelopes}.

Soft X-ray detectors have historically lagged behind technology for other wavelengths due to the inherent low signal levels associated with them and vacuum requirements in laboratory testing. For astronomy and astrophysics in particular, few commercial offerings exist that are suitable in cost and design flexibility for smaller scale missions.

In recent years, advancements in scientific Complimentary Metal Oxide Semiconductor (sCMOS), especially backside illumination, have enabled these sensors to become attractive detector candidates for X-ray missions. These options could theoretically be available at significantly lower cost and have lower power and cooling requirements, which could generate significant cost savings for space missions over CCDs.  Power and cooling are not particularly important concerns for a sounding rocket mission given the short duration (roughly 15 minute flight with 5 minutes of observation time).  However, for a future orbital version of this polarimeter, sCMOS detectors could bring significant cost savings.  Keeping this in mind we evaluated both CCD and sCMOS options for the REDSoX detectors.

One struggle in purchasing a camera from a commercial vendor for use in REDSoX, was the lack of testing data available in our band for many of the available sensors.  Soft X-ray facilities designed for low light levels are few and far between, and many of the commercial applications of these cameras are in the higher energies more commonly used in synchrotrons.  QE values in our bandpass are often calculated or extrapolated from higher energy data.  In order to be certain our requirements would be met by a candidate detector, it was necessary for us to carry out testing ourselves in our own soft X-ray facility.  Several candidate detector vendors provided us with demonstration model cameras that either contained the same chip we were evaluating, or one that was similar enough that we could be reasonably certain the chip proposed for our project would also fulfill our requirements.

Once we were satisfied that a given chip would meet our QE, spectral resolution, timing, and noise requirements, vendors provided us with concept designs to fit within the size requirements we provided, as well as quotes and timelines for delivery.

\section{MIT Polarimetry Beamline}

\begin{figure}
   \centering
    \includegraphics[width=17cm]{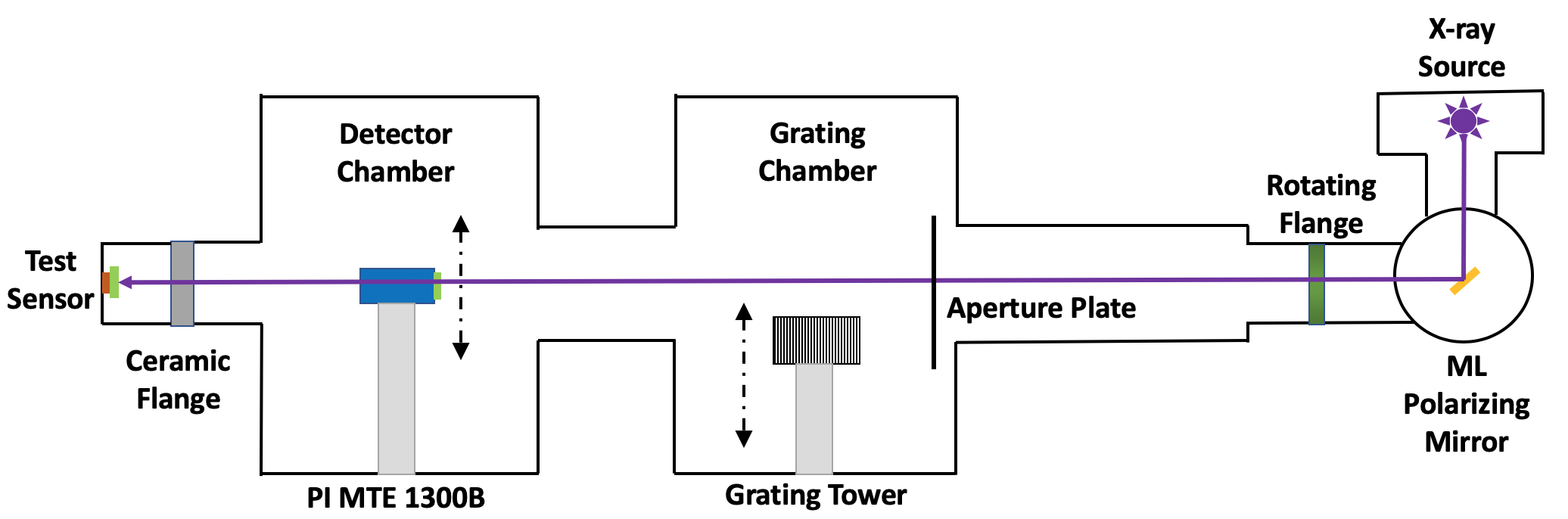}
 \caption{
\label{fig:Beamlinediagram}
A block diagram of the MIT polarimetry beamline.  The PI MTE 1300B detector is also installed on a motorized XY stage and can be inserted into the beam or moved out.  It is installed with an optical blocking filter as well as a shutter, which closes during readout.  There is a an electrically isolating ceramic flange on the end of the beamline, on which a flange-mounted detector or small chamber with a mounted sensor chip may be connected for testing.
}
\end{figure}

In order to evaluate candidate detectors for REDSoX, we installed each camera in the MIT polarimetry beamline\cite{murphy:77322Y,heine17,garner19}.  This facility has been re-purposed from its original use as a calibration facility for the Chandra HETG gratings and now serves as a testbed for REDSoX hardware and other technology development projects.  A block diagram of the beamline is shown in Figure \ref{fig:Beamlinediagram}.  

   
 The beamline is roughly 20 meters long and has three chambers (source, grating and detector).  The X-ray source is a Manson source with several interchangeable anodes to produce various line energies.  The source reflects off of a W/B$_4$C laterally graded multilayer mirror.  The mirror is a Bragg reflector, reflecting light polarized parallel to the mirror surface.  The lateral grading of the mirror means that the energy of the Bragg peak (energy of the reflected light) varies linearly along the surface of the mirror.  By selecting the anode material and the corresponding location to place the X-ray spot on the mirror, we reflect polarized, monochromatic light at the energy of our choice down the beamline.  We are able to tune the mirror to many emission lines between 183 eV and 700 eV without breaking vacuum.  The grating chamber contains an aperture plate, slit plate, and grating plate, which are all mounted on motorized stages.  Gratings installed on the grating plate can be inserted or removed from the beam using another motorized stage.

The detector chamber contains a Princeton Instruments MTE1300B CCD mounted on a motorized X-Y stage (perpendicular to the system optical axis).  Quick-look analysis is provided by custom Python-based code.  The motors and camera are controlled by custom Labview software.  There is also a flange on the end of the detector chamber allowing a test detector to be mounted in the beam path.

\vskip -0.1in
\vskip -0.1in


\section{Measurement Methodology}
\begin{figure}
   \centering
    \includegraphics[width=7cm]{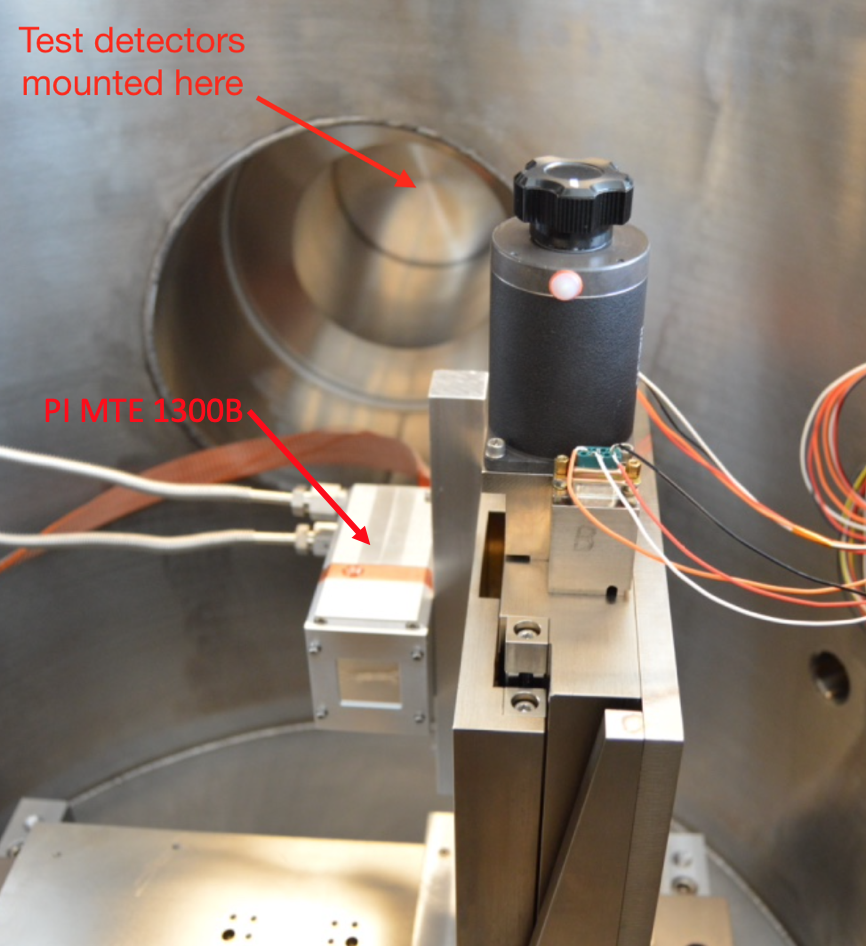}
 \caption{The inside of the detector chamber showing the mounted PI MTE1300B with an optical blocking filter installed as well as the flange onto which the test detectors are mounted.
}
\label{fig:Detectorchamber}
\end{figure}

We installed each camera for testing on an isolating flange on the end of the beamline.  This ceramic flange was necessary as we found that running a test detector without electrical isolation from the rest of the beamline led to excess readout noise.  Power for the detectors and their control and readout electronics was conditioned by a Tripp Lite power filter to further reduce any noise specific to the beamline itself or the building power.  We turned on a constant source current, which was monitored and recorded in case of fluctuations.  We utilized an open aperture allowing the full illumination of the detector in the beam path, then took data alternately with the Princeton Instruments MTE1300B and the test detector by moving the PI detector in and out of the beam path. A picture of the inside of the detector chamber is shown in Figure \ref{fig:Detectorchamber}.

Taking into account the extra distance to the test detector, we compared the count fluxes on both detectors to compare the QEs of the two detectors.  We also analyzed the energy resolution of each detector, along with uniformity across the chip, readout noise, and changes in these attributes with temperature and readout speed.  We obtained absolute QE measurements for each of the detectors by measuring their QEs relative to that of the PI MTE1300B.  We converted this to an absolute QE using the vendor-provided QE curve for the PI MTE detector, as modified by the optical blocking filter, purchased from Luxel.  We have not performed an absolute QE calibration of the MTE so it is possible that using the vendor QE curve may introduce systematic errors into our absolute QE measurements of the detectors under test.  However, for the purposes of selecting a detector for REDSoX we are satisfied that the QE of the PI detector is sufficient to meet the REDSoX requirements, so we can be assured that detectors that provide similar or better QE than the PI will meet the REDSoX requirements.

We tested four different detectors in the process of evaluating detectors for use in the REDSoX focal plane.  The cameras we evaluated were the Sydor Wraith, the Raptor Eagle XV, the Sony IMX290, and a CCID94 sensor produced and packaged by MIT Lincoln Labs.  The details of the chips utilized by these cameras are shown in Table 1, and the results of our testing and analysis are presented in the following sections.

\begin{table}
\label{tab:Sensors}
\begin{center}
\begin{tabular}{|p{3cm}|p{6cm}|p{3cm}|p{3.25cm}|}
\hline
Camera & Chip & Pixel Size ($\mu$m) & Pixel Number (Mpx) \\
\hline
Raptor Eagle XV & e2V 47-10 CCD & 13 & 4.2 \\
Sydor Wraith & Gpixel GSENSE 400 BSI sCMOS & 13 & 4.2 \\
Sony IMX290 & Sony IMX290 CMOS & 2.9 & 2.1 \\
MIT Lincoln Labs & CCID94 CCD & 24 & 2.1 \\
\hline
\end{tabular}
\caption{A summary of cameras/sensors tested.  The chip in the Raptor Eagle was about half the size required for REDSoX.  However the e2v 42-40 sensor  utilizes the same architecture and is twice the size, so if Raptor had been selected they would have built the REDSoX cameras with the larger chips.  The IMX290 chip is too small to meet the REDSoX requirements, but was included in our study to evaluate the potential of the Sony Starvis 1 series cameras for future soft X-ray applications.}
\end{center}
\end{table}

\section{Raptor Eagle XV}
\begin{figure}
   \centering
    \includegraphics[width=17cm]{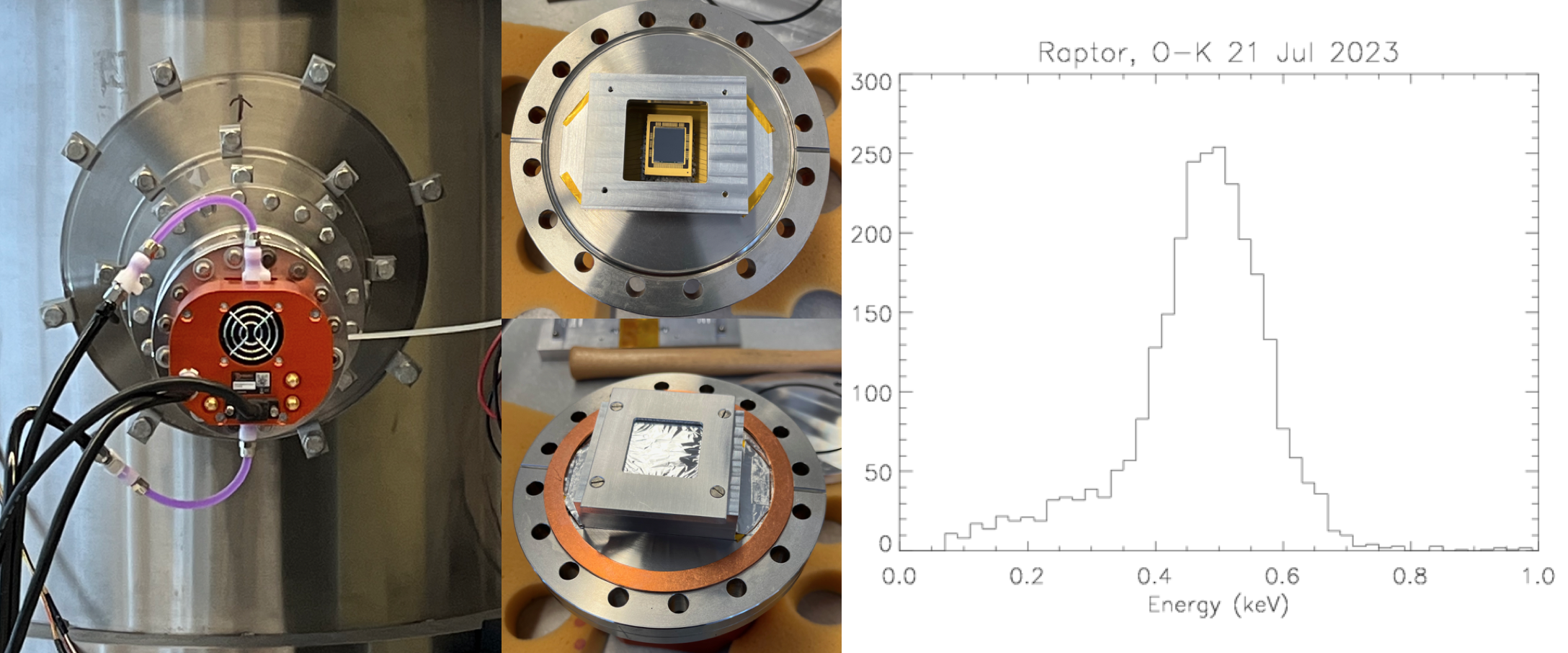}
 \caption{Left: a picture of the Raptor unit installed on the end of the polarimetry beamline.  Middle: (upper) the Raptor unit with the standoff installed without an optical blocking filter showing the bare CCD chip and (lower) the Raptor unit with optical blocking filter installed for testing.  Right: A pulse height distribution of the 525 eV oxygen-K line as measured by the Raptor Eagle demo unit in the beamline.
}
\label{fig:RaptorMounting}
\end{figure}

We tested a demo unit of the Raptor Eagle XV, which utilizes an e2V CCD47-10-1-355 chip.  This chip does not meet our size requirements, however it is similar enough to the CCD42-40 chip (which does meet our size requirements) that we were satisfied to test the smaller chip, then fly the larger chip if selected.  The Raptor demo unit did not have mounting holes for an optical blocking filter.  As our source produces enough optical light to cause noise issues on an unshielded CCD, we designed a standoff optical blocking filter mount that could be taped to the flange surface of the demo unit without damaging it.  This optical blocking filter mounting, as well as the mounting of the Raptor on the end of the beamline, are shown in Figure \ref{fig:RaptorMounting}.

We tested the Eagle at chip temperatures of -85 C and -50 C.  A ThermoCube chiller circulated coolant at 10 C to the hot side of a thermoelectric cooler (TEC) in the flange mounted Raptor camera package.  We also tested readout rates of 75 kHz and 2 MHz.  In (off-line) processing, we observed a variation in the count rate in the Y direction due to a frame wipe in the readout process, so we accounted for this effect in the final analysis.  A flight unit would be operated without the frame wipe option.

We first verified the gain value by looking at the peak DN for each of several known source energies.  Our next test was to verify QE.  We measured the QE in comparison with that of the e2V CCD36-40 chip in our PI MTE 1300B, which we are confident is sufficient to meet the requirements for a REDSoX detector.  In a single session utilizing the oxygen line, we observed with both the PI and the Raptor cameras.  After correcting for the extra distance to the Raptor and for the variation due to the frame wipe, we found the QE of the Raptor camera to be within 2\% of that of the PI camera.  

\begin{figure}
   \centering
    \includegraphics[width=17cm]{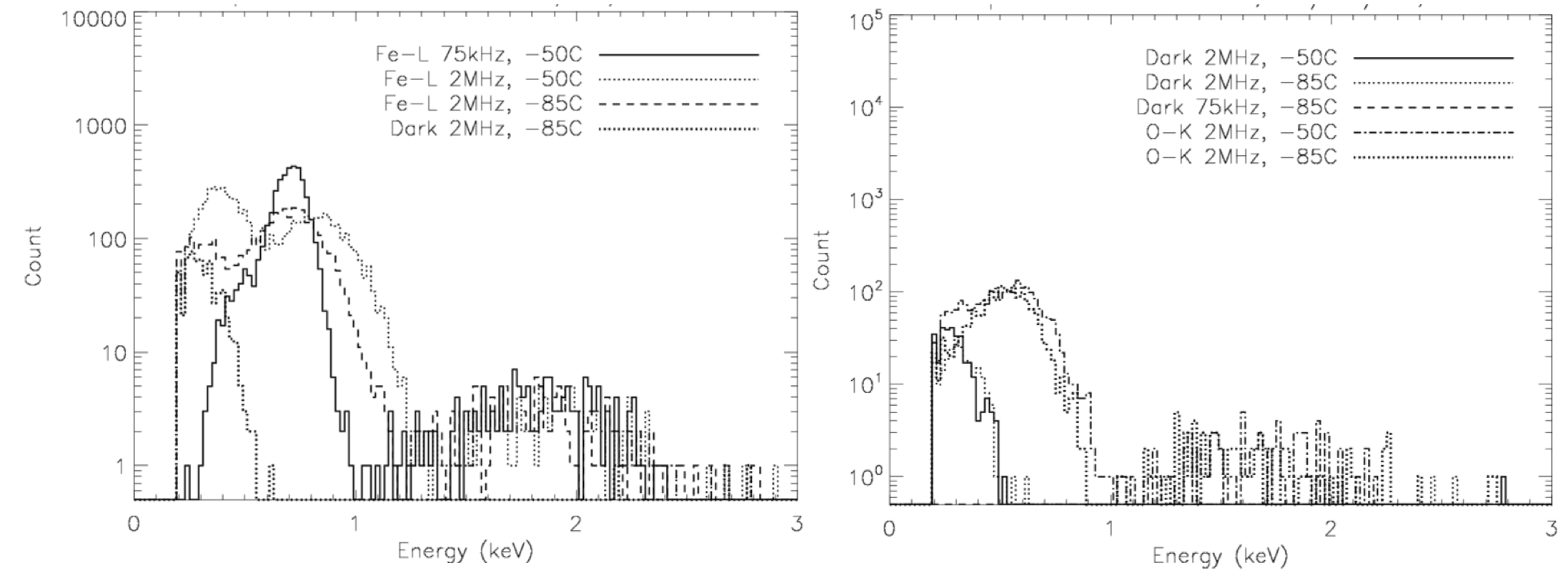}
 \caption{Pulse height distributions for Fe-L$\alpha$ 705 eV (left) and O-K$\alpha$ 525 eV (right) data measured with the Raptor Eagle in the beamline.  Emission in the 1.2-2.3 keV region is due to fluorescence lines of Al-K$\alpha$ (in the mirror holder) and W-M$\alpha$ (in the mirror's multilayer).
}
\label{fig:RaptorPHD}
\end{figure}

Our next tests of the Raptor were to verify the readout noise and spectral resolution.  For the readout noise we exposed the sensor to both the iron L$\alpha$ line at 705 eV, and the oxygen K$\alpha$ line at 525 eV, as well as taking dark frames for all combinations of chip temperatures of -50 C and -85 C and readout rates of 75 kHz and 2 MHz.  The pulse height distributions from these tests are shown in Figure \ref{fig:RaptorPHD}.  Our conclusions were that readout noise was sufficiently low at -50 C for our purposes.  The 2 MHz readout rate noise was sufficient for our imaging camera, but would not be sufficient for the polarimetry cameras.

\begin{figure}
   \centering
    \includegraphics[width=17cm]{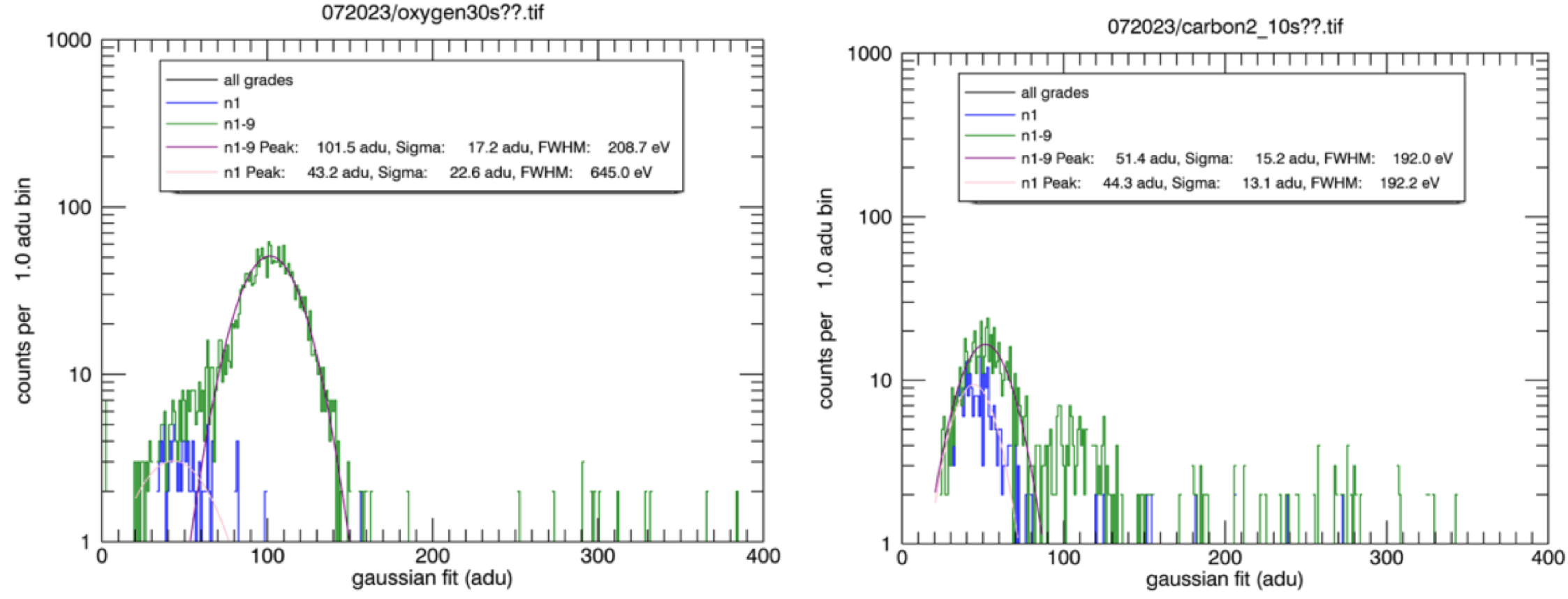}
 \caption{Fits to O-K$\alpha$ 525 eV (left) and C-K$\alpha$ 277 eV (right) data to measure spectral resolution of the Raptor Eagle chip in the beamline.
}
\label{fig:RaptorSpectralRes}
\end{figure}

We also measured the spectral resolution of the chip using both oxygen and carbon data.  The data and fits from these tests are shown in Figure \ref{fig:RaptorSpectralRes}.  Using all grades, the FWHM measured for the oxygen line was 209 eV while the FWHM measured for the carbon line was 192 eV despite the distribution being clipped at low energy.

As a result of these tests we determined that the Raptor Eagle met the REDSoX detector requirements.  Raptor produced a mechanical concept and quote for flight detectors, however we did not select Raptor as our detector vendor for programmatic reasons.

\section{Sydor Wraith}
\begin{figure}
   \centering
    \includegraphics[width=7cm]{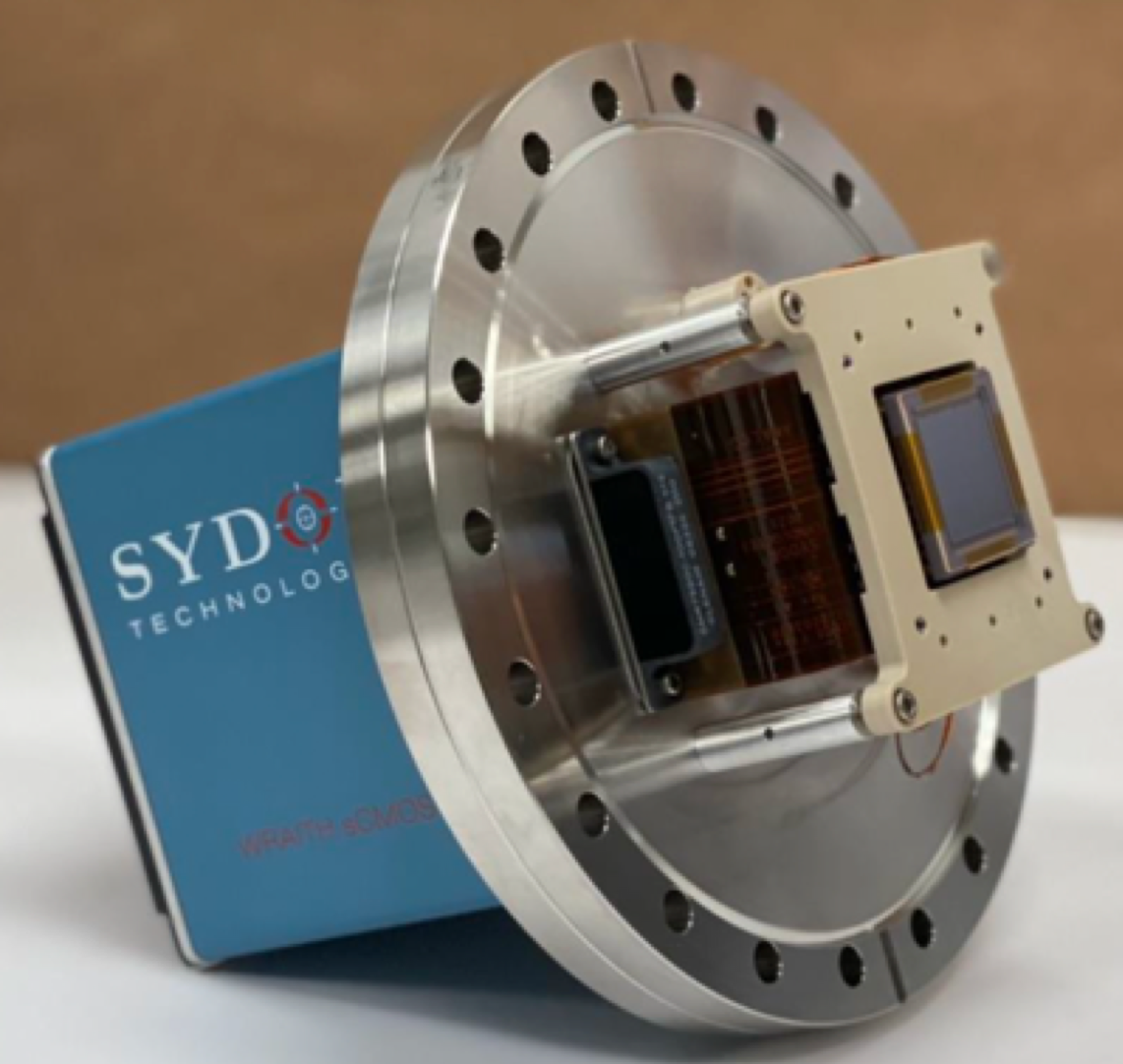}
 \caption{The Sydor Wraith detector with a glass slide taped over the bare CMOS sensor for protection.  The glass slide was removed before installing on the beamline.
}
\label{fig:Sydorimage}
\end{figure}

\begin{figure}
   \centering
    \includegraphics[width=9cm]{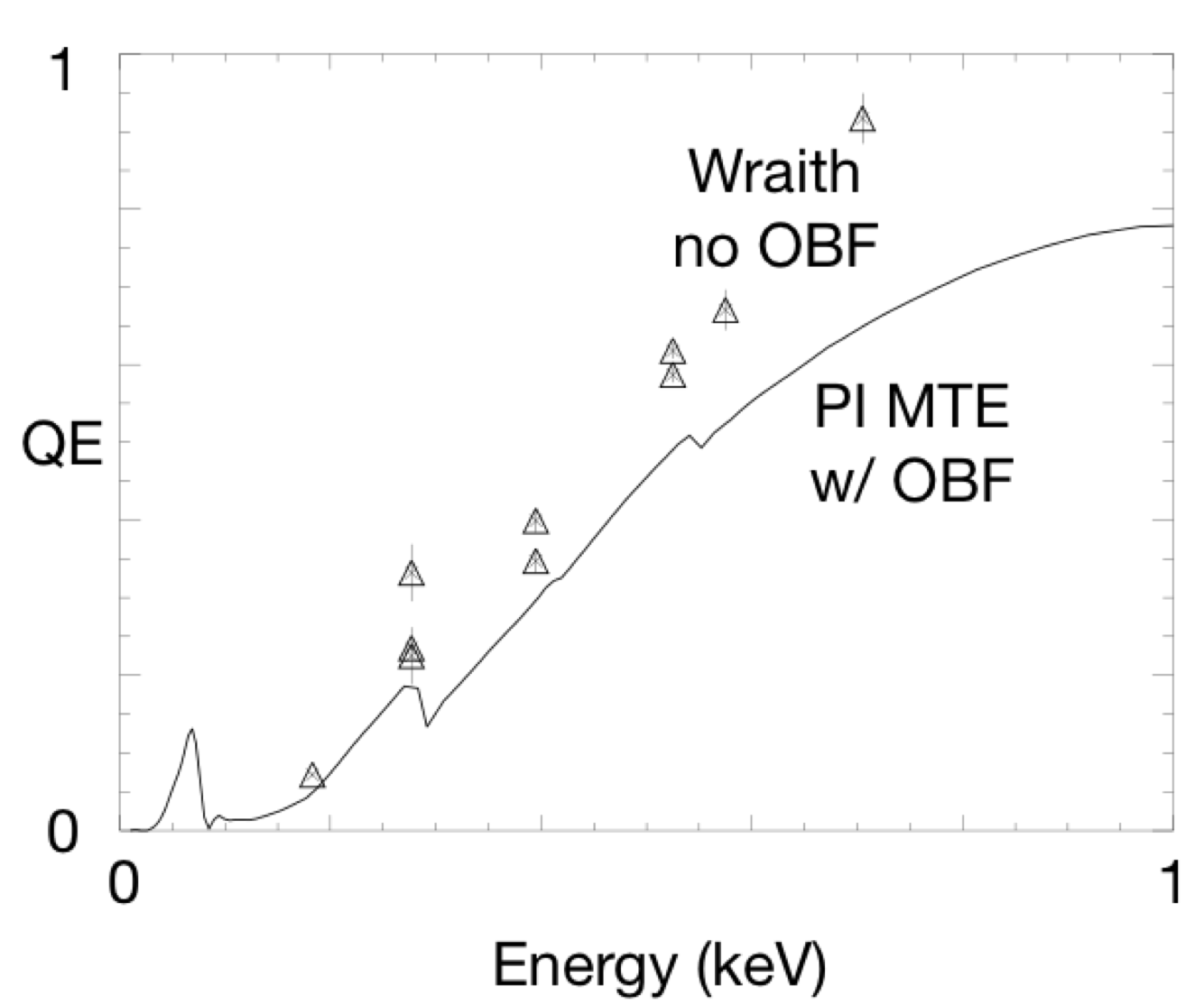}
 \caption{The measured QE of the Sydor Wraith with no optical blocking filter (run with 100 ms frames to allow us to take out the optical contamination as noise) vs the QE of the PI MTE with an optical blocking filter.  The PI curve is obtained by multiplying the estimated QE curve of its CCD by the throughput vs.\ energy of the OBF as calculated using Henke optical constants\cite{henke93}.  
}
\label{fig:SydorQE}
\end{figure}

We utilized the same testing methodology with the Sydor Wraith, which is a flange-mounted camera utilizing the Gpixel GSENSE 400BSI chip, an sCMOS sensor.  A picture of the Sydor Wraith is shown in Figure \ref{fig:Sydorimage}.  We originally tested the Sydor Wraith as part of an SBIR grant partnering with Sydor.  During that investigation we had installed an optical blocking filter on the camera and had measured a very low QE (about a third to a half of what would be required for REDSoX).  However, in our subsequent experience with sCMOS sensors we found that operating quickly enough (roughly 100 ms frame times), we were able to collect enough frames without X-rays on all pixels to remove the optical contamination as a baseline.  Thus, when running quickly enough, which can only be done with a CMOS sensor rather than a CCD, we can run the detector without an optical blocking filter, raising the effective QE of the detector.  A plot of the QE of the bare Sydor camera compared to the QE of the PI MTE with an optical blocking filter is shown in Figure \ref{fig:SydorQE}.  We tested the performance of the Wraith at -25, -15, and 0 C.  The sensor was cooled by a TEC, the hot side of which was chilled with 10 C coolant circulated by a Thermocube chiller.

\begin{figure}
   \centering
    \includegraphics[width=17cm]{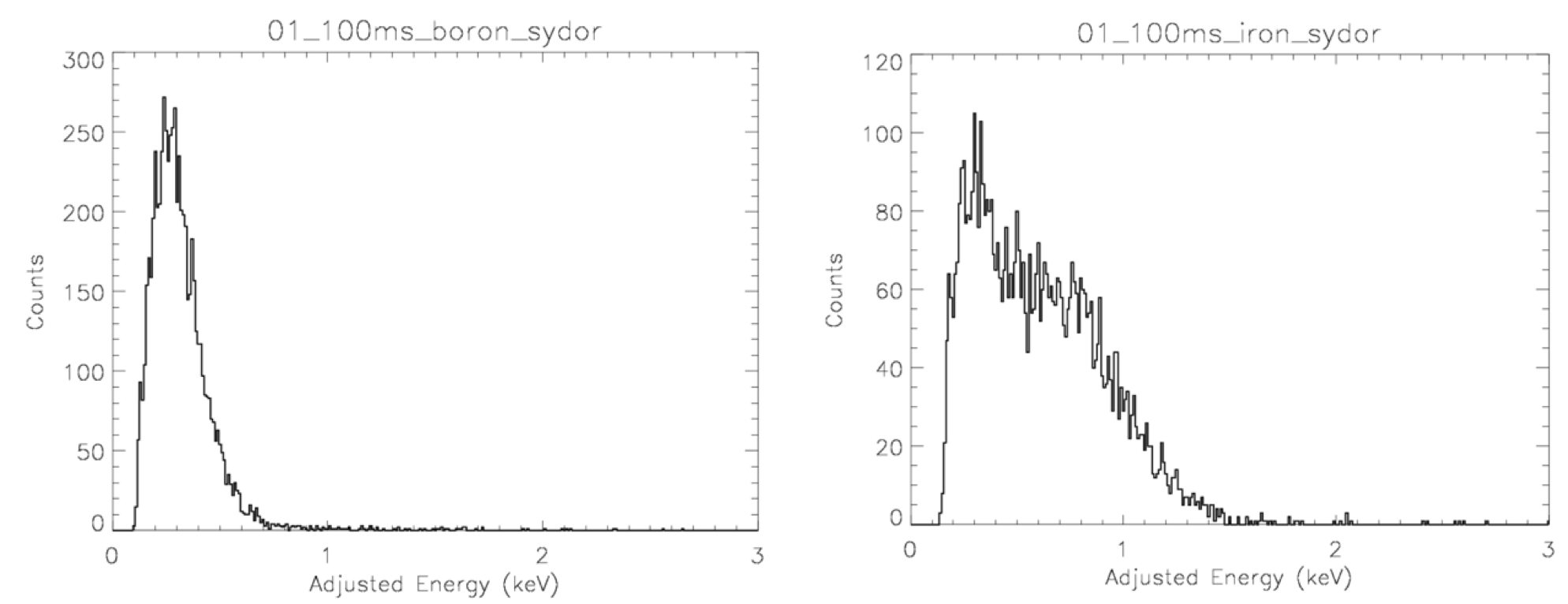}
 \caption{Pulse height distributions for B-K$\alpha$ (left) and Fe-L$\alpha$ (right) measured with the Sydor Wraith.
}
\label{fig:SydorPHD}
\end{figure}

We observed more intensive charge spreading than expected in the events measured with the Wraith.  To retrieve the full energy of measured events, we found we needed to use 5$\times$5 or 7$\times$7 event islands for event extraction.  We also noted a significant difference in the noise and event rates of the odd versus the even rows.  We were not able to discern if this was native to the chip itself or was an issue caused by readout, but we were able to mitigate the effect with 2$\times$2 binning.  The pulse height distributions for both boron K$\alpha$ and iron L$\alpha$ are shown in Figure \ref{fig:SydorPHD}.  Based on the results of our testing we determined that the Sydor Wraith meets the REDSoX detector requirements.

Sydor produced a mechanical concept for a flight detector based largely on the boards and configuration utilized inside the body of the flange mounted detector.  However, the fit of this initial mechanical configuration was extremely tight in the focal plane.  After studying the layout of the boards it became clear that modifying them to fit in the case that we needed to change the layout of our focal plane in any way would be cost prohibitive.  Due to this risk, we did not select Sydor as the detector vendor for REDSoX.

\vskip -0.1in
\section{Sony IMX290}
Another CMOS option we explored was to use inexpensive, commercial off the shelf (COTS) CMOS packages designed for use in the optical regime with modifications to make them usable in the X-ray band.  The IMX290 from the Sony Starvis series has been investigated for use in the X-ray band elsewhere \cite{TammesIMX290}.  Despite its size being insufficient to meet the REDSoX requirements, we tested this sensor for the purpose of determining whether this class of sensor could be a viable option for REDSoX.  Schneider et al \cite{IMX290perpixel} provide more complete information on testing of the IMX290 sensor at MIT, including characterizations beyond those performed in the polarimetry beamline.

\begin{figure}
   \centering
    \includegraphics[width=15cm]{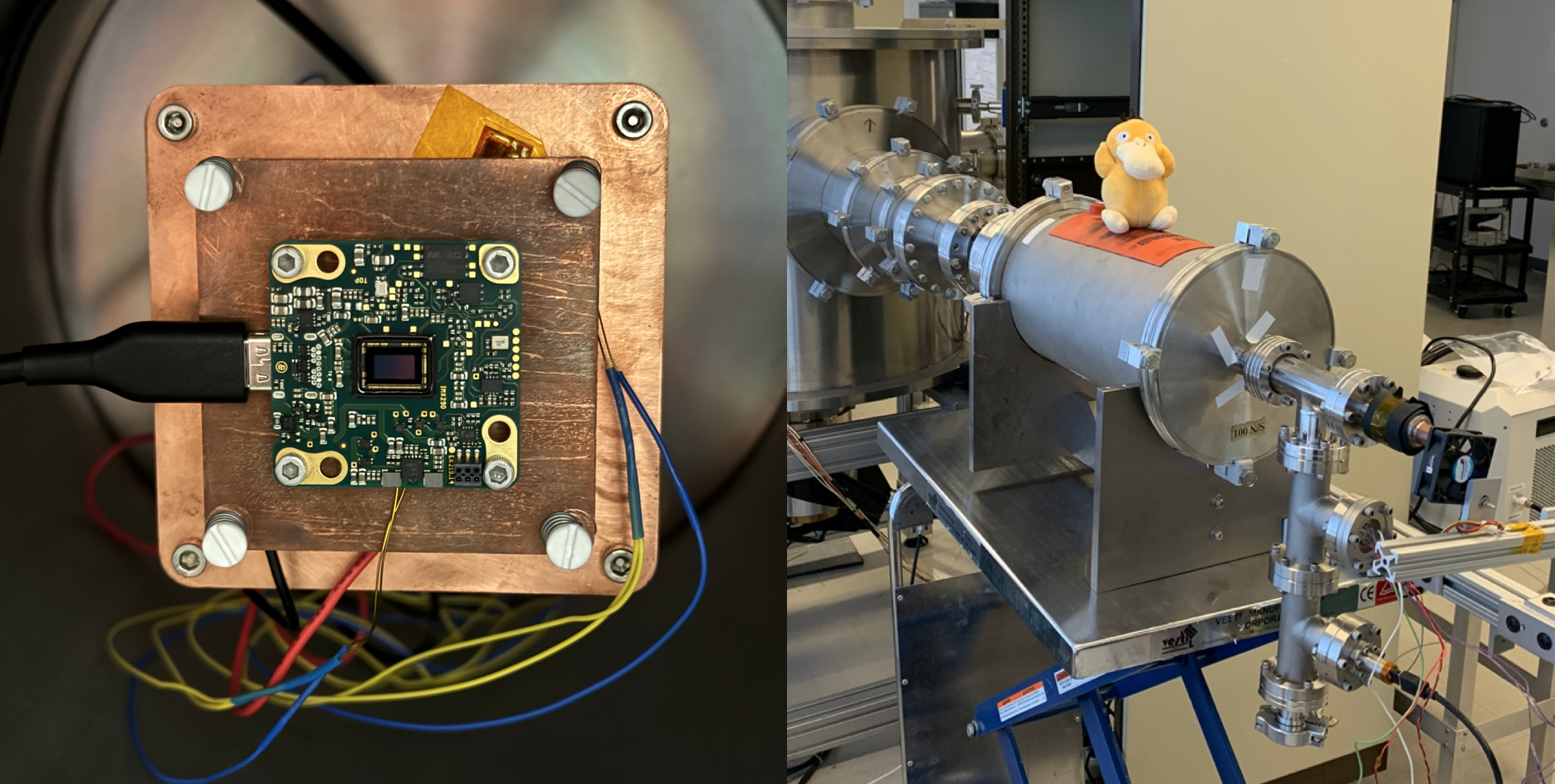}
 \caption{The experimental setup for testing the IMX290. Left: the setup of the CMOS sensor inside its vacuum chamber.  The sensor is visible at the center of the image and surrounded by its readout electronics. The camera is connected to the cold side of the TEC (not visible) via a copper plate. The hot and cold sides are held together with PTFE plastic screws to minimize the heating load.  Right: the IMX290 vacuum chamber attached to the end of the polarimetry beamline for testing.
}
\label{fig:Psyducksetup}
\end{figure}

The chips are sold with a glass cover for use with optical light.  We had this glass removed by an outside vendor to allow X-rays to get to the sensor.  The camera was attached to the cold side of a thermoelectric cooler via a copper plate as shown in Figure \ref{fig:Psyducksetup}.  As with the Sydor Wraith, we were able to run the IMX290 at 100 ms or shorter frames, enabling us to remove the optical light as noise so we did not utilize an optical blocking filter on the sensor.  The hot side of the thermoelectric cooler was attached to a metal plate, which was cooled outside of vacuum with a fan.  The main purpose of the TEC was to keep the temperature of the sensor stable.  We operated the chip at temperatures of 25-30 C, much warmer than the temperatures used for any of the other candidate detectors.  A picture of the IMX290 vacuum chamber on the beamline with the TEC hotside and electronics feedthroughs attached is shown in Figure \ref{fig:Psyducksetup}.

\begin{figure}
   \centering
    \includegraphics[width=17cm]{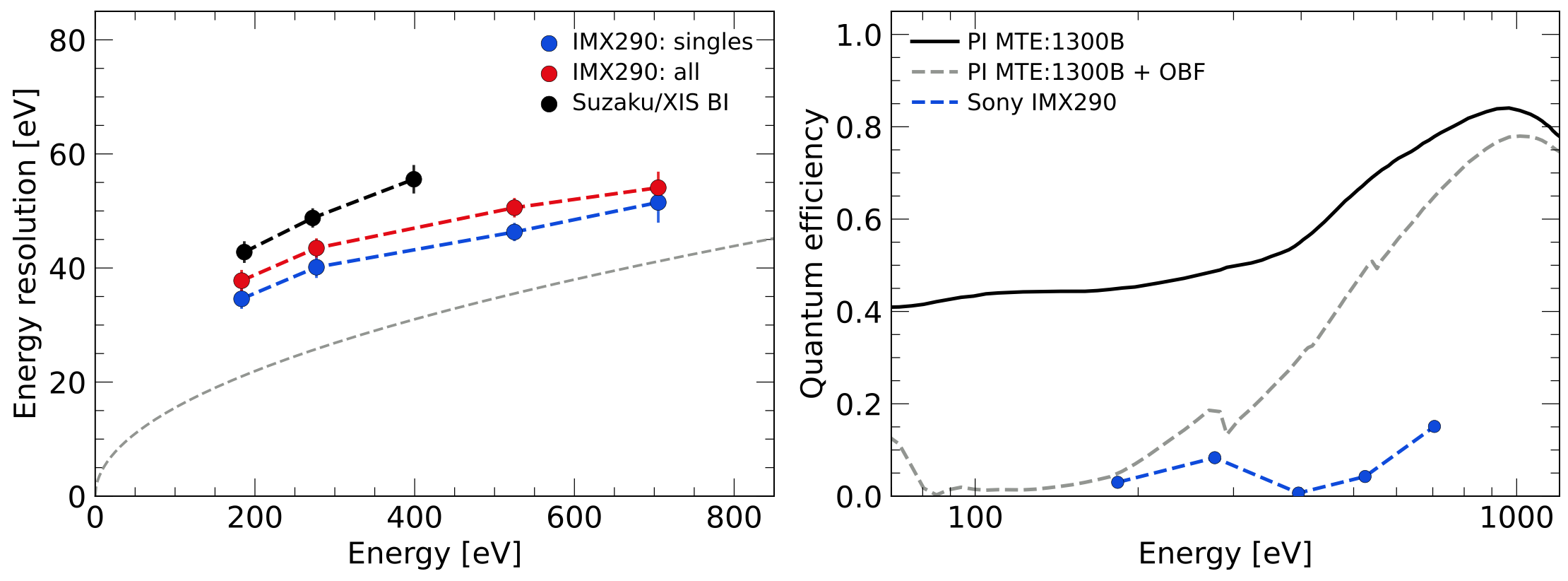}
 \caption{Left: The energy resolution obtained for several different event energies utilizing the IMX290.  The energy resolution from events falling completely within a single pixel are shown in blue while the energy resolution for all events is shown in red.  The fano limit is shown as a gray dashed line.  The energy resolution of the Suzaku XIS detectors is shown in black for comparison, demonstrating that the IMX290 is able to measure lines with better energy resolution than the Suzaku CCDs.  Right: Measured QE of the IMX290 compared with the bare MTE1300B and the MTE1300B with an optical blocking filter.  We believe the poor QE of the IMX290 is a result of a microlens array, likely made of some kind of plastic, on the surface of the chip.
}
\label{fig:Psyduckplots}
\end{figure}

Our testing revealed very impressive energy resolution results for the IMX290, approaching the Fano limit.  The results of our energy resolution tests are shown in Figure \ref{fig:Psyduckplots}.  Unfortunately the measured QE of the sensor was quite low (much worse than that of the MTE1300B with an optical blocking filter).  We subsequently confirmed with the vendor of the Sony IMX290 that there is a microlens array on the surface of the chip that likely caused this poor QE by absorbing a large fraction of the soft X-rays.  While we were able to find outside vendors who would attempt to remove the microlens array in addition to the cover glass, this process is still in development, and it is not guaranteed the sensor will remain undamaged.  Due to this production uncertainty we decided not to move forward with Starvis series chips for use in REDSoX.  However, the impressive energy resolution, low sensor cost, available high readout speeds, and ability to run the sensor very warm are extremely promising, so we are continuing to investigate these sensors for use in future missions.
\vskip -0.1in
\section{MITLL CCID94}
We also evaluated the CCID94 CCD sensor produced by MIT Lincoln Laboratories as a candidate for a REDSoX sensor.  More detailed information about this sensor can be found in a paper by Miller et al \cite{Miller2024_SPIE}.  These chips provide an attractive option because of the long heritage and high level of expertise in creating high quality CCDs for X-ray astrophysics missions at MIT Lincoln Lab, as well as the expertise and experience in working with these chips in the X-ray detector lab at the MIT Kavli Institute.  There was also a possibility that spare readout electronics from the REXIS project, which utilized very similar chips from MITLL, could be used for the readout for REDSoX.  Unfortunately it was unclear if the REXIS electronics could provide the necessary readout noise to meet REDSoX requirements without modification.

\begin{figure}
   \centering
    \includegraphics[width=12cm]{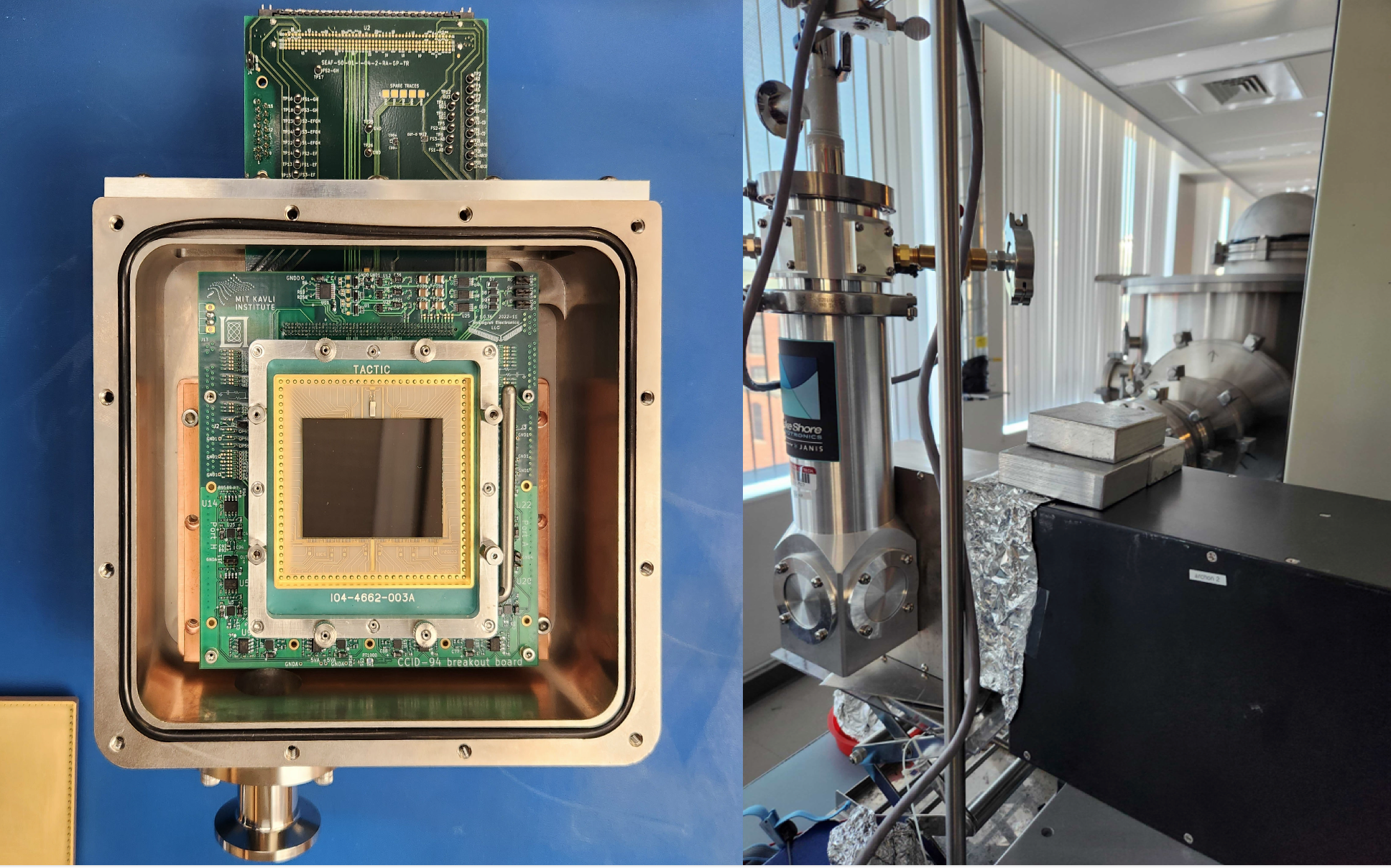}
 \caption{Left: The packaged CCID94 chip from MITLL.  Right: the vacuum chamber containing the CCID94 and its associated laboratory readout electronics attached to the end of the beamline.
}
\label{fig:CCID94Setup}
\end{figure}

The chips were fabricated and packaged at MITLL.  The CCD was cooled using liquid nitrogen, with flow controlled by a Lake Shore temperature control box.  The packaged sensor and the beamline setup are shown in Figure \ref{fig:CCID94Setup}.  The packaging did not include mounting holes for an optical blocking filter.  Despite shorter frame times than we typically use for our PI camera (we used frame times of 1 second for the CCID94) optical contamination from our Manson source created excessive noise that interfered with our ability to extract X-ray events cleanly from the data.  We mitigated this problem by installing an optical blocking filter on a motorized stage in the grating chamber.  We were able to move the filter into the beam path when we were taking data on the CCID94, and out of the beam path when we were taking data with the PI MTE1300B, which has an identical optical blocking filter installed on its front.  This approach effectively eliminated optical light during data acquisition.

\begin{figure}
   \centering
    \includegraphics[width=11cm]{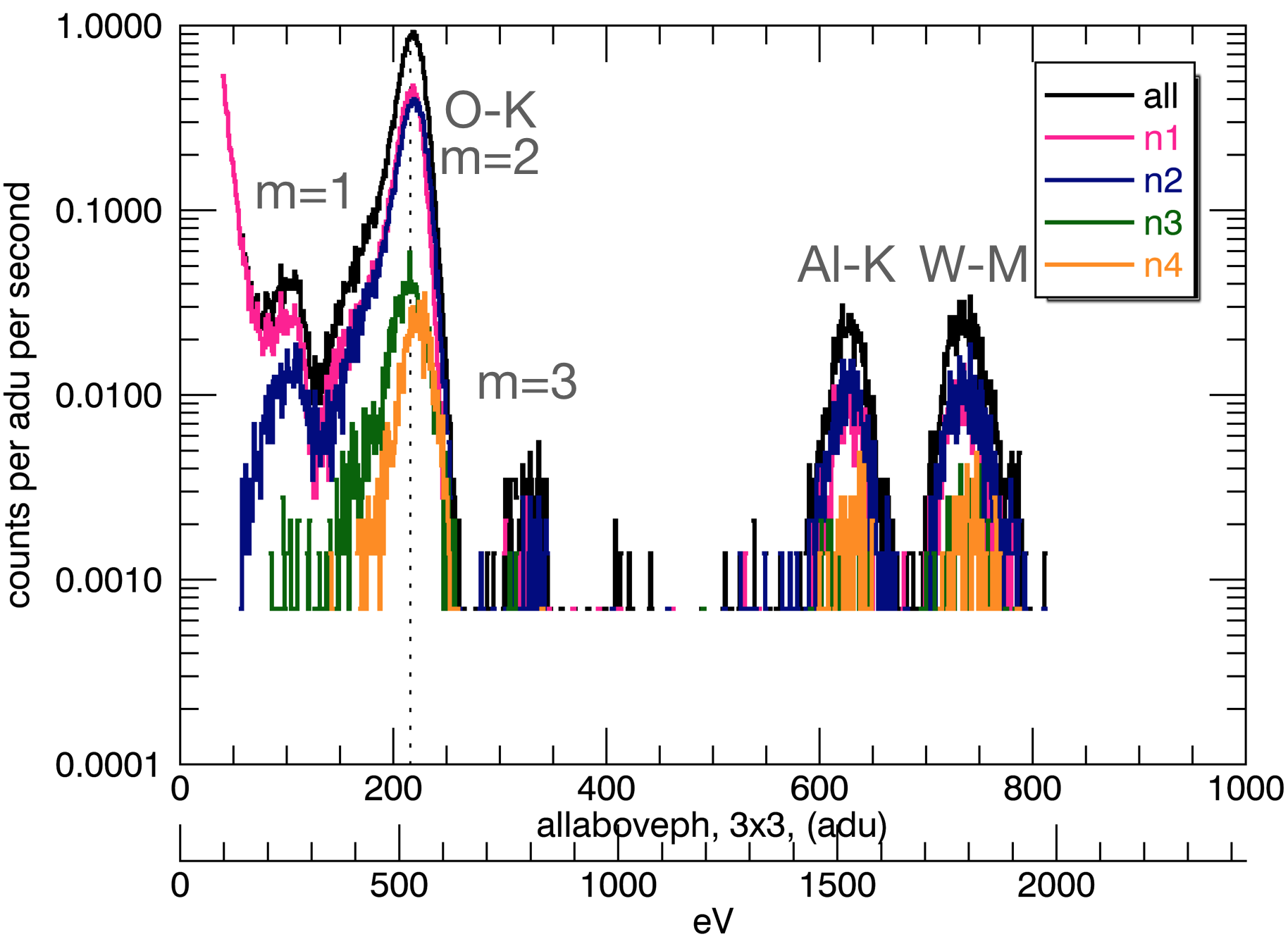}
 \caption{Pulse height distribution of data taken with a Sapphire anode to produce an O-K$\alpha$ line.  Three orders of the line reflected by the multilayer mirror were detected.  Also detected are Al-K$\alpha$ fluorescence from the mirror holder and W-M$\alpha$ fluorescence from the mirror's multilayer coating (compare to Fig.~\ref{fig:RaptorPHD}right).
}
\label{fig:CCID94PHD}
\end{figure}

A pulse height distribution from measurement of O-K$\alpha$ in the beamline is shown in Figure \ref{fig:CCID94PHD}.  We measured very good spectral resolution of 53 eV for single pixel events and 57 eV for all events for the oxygen line.  While it was a very attractive candidate, we did not select the CCID94 sensor for our REDSoX detectors due to programmatic reasons.

\section{Future Work}

After a very extensive process testing candidate detectors and corresponding with detector vendors we selected XCAM as our detector vendor.  They will be producing cameras for our project using the e2V CCD207-40 chip.  After studying the details of this chip we deemed it to be substantially similar to the e2V CCD47-10 we tested while evaluating the Raptor Eagle, and decided we did not need to test this chip before proceeding with purchase.  XCAM has facilities to test the cameras to ensure they meet our requirements before delivery.  We expect delivery of an engineering unit in the fall of 2024.  Once we have received the engineering unit we will be able to proceed with testing of the unit itself, in addition to testing and development of interface cabling and control and readout software.  Delivery of the flight units will follow in the following year.  The REDSoX launch is scheduled for summer of 2027 from White Sand Missile Range in New Mexico.

 Our group is also continuing to test detectors, especially sCMOS detectors, with the goal of identifying prospective detectors for a future orbital polarimeter mission.  We are currently evaluating the PCO/Excelitas Edge 4.2 BI utilizing a Gpixel 2020BSI chip, as well as beginning work with the Sony IMX226, with both the cover glass and lenslet array removed.  

\bibliography{polarimetry20}   

\begin{thebibliography}{10}

\bibitem{redsoxjatis}
{Marshall}, H.~L., {G{\"u}nther}, H.~M., {Heilmann}, R.~K., {Schulz}, N.~S., {Egan}, M., {Hellickson}, T., {Heine}, S.~N.~T., {Windt}, D.~L., {Gullikson}, E.~M., {Ramsey}, B.~D., {Tagliaferri}, G., and {Pareschi}, G., ``{Design of a Broad-band Soft X-ray Polarimeter},'' {\em Journal of Astronomical Telescopes, Instruments, and Systems}~{\bf 4},  11004 (Mar. 2018).

\bibitem{redsox24}
{Garner}, A., {Marshall}, H.~L., {Heine}, S. N.~T., {Schulz}, N.~S., {Heilmann}, R.~K., {Gunther}, H.~M., {Juneau}, J., {LaMarr}, B., {Metivier}, A., {Ravi}, S., {Kothnur}, N.~V., {Bongiorno}, S.~D., and {Gullikson}, E.~M., ``Current status of the redsox sounding rocket, in prep,'' in [{\em Society of Photo-Optical Instrumentation Engineers (SPIE) Conference Series}{\nolinebreak\hspace{0.1em}]},  {\em Society of Photo-Optical Instrumentation Engineers (SPIE) Conference Series} {\bf 13093} (2024).

\bibitem{2022HeilmannCAT}
{Heilmann}, R.~K., {Bruccoleri}, A.~R., {Burwitz}, V., {DeRoo}, C., {Garner}, A., {Gunther}, H.~M., {Gullikson}, E.~M., {Hartner}, G., {Hertz}, E., {Langmeier}, A., {Muller}, T., {Rukdee}, S., {Schmidt}, T., {Smith}, R.~K., and {Schattenburg}, M.~L., ``{X-ray Performance of Critical-angle Transmission Grating Prototypes for the Arcus Mission},'' {\em \apj}~{\bf 934},  15pp (Aug. 2022).

\bibitem{murphy:77322Y}
{Murphy}, K.~D., {Marshall}, H.~L., {Schulz}, N.~S., {Jenks}, K., {Sommer}, S. J.~B., and {Marshall}, E.~A., ``{Soft x-ray polarimeter laboratory tests},'' in [{\em Space Telescopes and Instrumentation 2010: Ultraviolet to Gamma Ray}{\nolinebreak\hspace{0.1em}]},  {\em Society of Photo-Optical Instrumentation Engineers (SPIE) Conference Series} {\bf 7732},  77322Y (July 2010).

\bibitem{heine17}
{Heine}, S.~N.~T., {Marshall}, H.~L., {Heilmann}, R.~K., {Schulz}, N.~S., {Beeks}, K., {Drake}, F., {Gaines}, D., {Levey}, S., {Windt}, D.~L., and {Gullikson}, E.~M., ``{Laboratory progress in soft x-ray polarimetry},'' in [{\em Society of Photo-Optical Instrumentation Engineers (SPIE) Conference Series}{\nolinebreak\hspace{0.1em}]},  {\em Society of Photo-Optical Instrumentation Engineers (SPIE) Conference Series} {\bf 10399},  1039916 (Aug. 2017).

\bibitem{garner19}
{Garner}, A., {Marshall}, H., {Heine}, S., {Heilmann}, R., {Song}, J., {Schulz}, N., {LaMarr}, B., and {Egan}, M., ``{Component testing for x-ray spectroscopy and polarimetry},'' in [{\em Society of Photo-Optical Instrumentation Engineers (SPIE) Conference Series}{\nolinebreak\hspace{0.1em}]},  {\em \procspie} {\bf 11118},  1111811--1--12 (Sept. 2019).

\bibitem{henke93}
{Henke}, B.~L., {Gullikson}, E.~M., and {Davis}, J.~C., ``{X-Ray Interactions: Photoabsorption, Scattering, Transmission, and Reflection at E = 50-30,000 eV, Z = 1-92},'' {\em Atomic Data and Nuclear Data Tables}~{\bf 54},  181--+ (1993).

\bibitem{TammesIMX290}
{Tammes}, S., {Roth}, T., {Kaaret}, P., {DeRoo}, C., {Elmaleh}, A., {McChesney}, J., and {Rodalakis}, F., ``Soft x-ray detection for small satellites with a commercial cmos sensor at room temperature,'' {\em JATIS}~{\bf 6} (2020).

\bibitem{IMX290perpixel}
{Schneider}, B., {Prigozhin}, G., {Foster}, R.~F., {Bautz}, M.~W., {Fu}, H., {Grant}, C.~E., {Heine}, S., {Juneau}, J., {LaMarr}, B., {Limousin}, O., {Lourie}, N., {Malonis}, A., and {Miller}, E.~D., ``X-ray spectral performance of the sony imx290 cmos sensor near fano limit after a per-pixel gain calibration, in prep,'' {\em JATIS}  (2024).

\bibitem{Miller2024_SPIE}
Miller, E.~D., Gregory, J.~A., Bautz, M.~W., Clark, H.~R., Cooper, M.~J., Donlon, K., Foster, R.~F., Grant, C.~E., Jensen, M., LaMarr, B.~J., Lambert, R.~D., Leitz, C., Malonis, A., Neak, M., Prigozhin, G., Ryu, K.~K., Schneider, B., Warner, K., Young, D.~J., and Zhang, W.~W., ``{Curved detectors for future X-ray astrophysics missions},'' in [{\em Space Telescopes and Instrumentation 2024: Ultraviolet to Gamma Ray}{\nolinebreak\hspace{0.1em}]},  den Herder, J.-W.~A., Nikzad, S., and Nakazawa, K., eds.,  {\bf 13093},  13093--217, International Society for Optics and Photonics, SPIE (2024).

\end{thebibliography}
\bibliographystyle{spiebib}   
\include{bibliography}
\end{document}